\documentclass[%
 reprint,
 superscriptaddress,
 amsmath,amssymb,
 aps,
 pra,
]{revtex4-2}

\usepackage{graphicx}
\usepackage{dcolumn}
\usepackage{bm}
\usepackage{enumerate}
\usepackage{braket}
\usepackage{xcolor}

\usepackage{xcolor}

\begin{document}

{\title{Experimental observation of quantum interferences in CO-H$_2$ rotational energy transfer at room temperature}

\author{Hamza Labiad}
\email[]{hamza.labiad@stfc.ac.uk} 
\affiliation{Univ Rennes, CNRS, IPR (Institut de Physique de Rennes) - UMR 6251, F-35000 Rennes, France}

\author{Alexandre Faure}
\email[]{alexandre.faure@univ-grenoble-alpes.fr} 
\affiliation{Universit\'e de Grenoble Alpes, CNRS, IPAG, F-38000 Grenoble, France}

\author{Ian R. Sims}
\email[]{ian.sims@univ-rennes.fr}
\affiliation{Univ Rennes, CNRS, IPR (Institut de Physique de Rennes) - UMR 6251, F-35000 Rennes, France}
\affiliation{Institut universitaire de France (IUF)}

\begin{abstract}
Using time-resolved infrared–vacuum-ultraviolet double-resonance spectroscopy, experimental room temperature measurements of state-to-state rate coefficients for rotational energy transfer within the X $^1\Sigma^+(v=2)$ vibrational state of CO due to H$_2$ collisions have been compared to accurate 4-D close-coupling quantum calculations. Theoretically predicted quantum interferences in the CO-H$_2$ collisional system are experimentally observed at room temperature, and excellent agreement between theory and experiment is observed. These results provide a valuable benchmark for validating the anisotropic part of the potential energy surface, thereby supporting the theoretical modeling of CO emission in warm astrophysical environments such as photodissociation regions.
  
\end{abstract}

\maketitle

\section{Introduction}
Interacting systems inducing collisional energy transfer, are of great interest in many areas, ranging from astrochemistry \cite{roueff_molecular_2013, faure_collisional_2022}, atmospheric science \cite{Noll_atmosRET_2018}, plasma and combustion \cite{vorac_state-by-state_2017, jorg_state-specific_1992, jo_rovibrational_2023}, and gas lasers \cite{ratanavis_performance_2009, nampoothiri_hollow-core_2012, song_218_2024, shi_mid-infrared_2024} to fundamental physical chemistry \cite{bernstein_atom_1979}. Measuring collision induced rotational energy transfer, RET, becomes crucial especially when chemical and physical processes of non-equilibrium gas phase environments are involved. Rate coefficients for RET are very sensitive to the potential energy surface (PES) of inelastic colliding systems making their measurement an ideal tool to accurately benchmark the quantum mechanical calculations with the objective to better understand the quantum phenomena that govern the elementary physico-chemical processes.\\
Measuring RET rate coefficients at very low temperatures ($T \lesssim 100$~K) has received a significant interest in the last decades because of the need for these rate coefficients in the fast-growing field of astrochemistry, and because of many interesting  phenomena which are purely of a quantum nature, such as shape resonances. Experimental techniques have been significantly pushed forward for this purpose \cite{labiad_absolute_2022, chefdeville_appearance_2012}, of course in addition to other cold chemistry purposes.\\
Somewhat less attention has been paid to the problem of measuring RET rate coefficients at relatively high temperatures despite its importance in completing the bigger picture of the theoretical benchmarking of both the attractive and repulsive parts of the PES. This is in part because techniques for accurately measuring RET rate coefficients have mainly been developed and adapted for low temperature studies. \\
On the other hand, even though diatom-diatom collisional systems may still appear relatively simple, they represent the most basic and important systems that must be well understood to gain confidence in studying more complex polyatomic collisional systems. In fact, very accurate full dimensional quantum close-coupling calculations on diatom-diatom systems were not possible until very recently. The quantum calculations of these systems become more complex if the number of channels is large (as is the case at higher temperatures for example), and even more complex if the collisional partner is an open-shell molecule. Recent examples include the CS + H$_2$ \cite{Yang2018}, SO + H$_2$ \cite{Yang2020} and HCl + H$_2$ \cite{Hoffman2024} systems. \\
The CO-H$_2$ system is an ideal diatom-diatom prototype for collisional systems involving nuclear-spin wave function of H$_2$, and a nearly homonuclear type molecule, CO. Interestingly, studying such a system can exhibit quantum effects at even relatively high temperatures in addition to low temperature quantum resonance effects, as observed in the CO-He \cite{bergeat_quantum_2015}, or in the NO-H$_2$ \cite{vogels_scattering_2018} systems. In fact, the quantum interference effect, predicted theoretically by the semiclassical model calculations of McCurdy and Miller \cite{mccurdy_interference_1977}, can be observed more easily at higher temperatures thanks to more transitions being available compared to lower temperatures. These interferences are characterized by strong propensity rules in final state populations. They are a consequence of the wave nature of matter and cannot be properly described by a quasi-classical trajectory (QCT) theory. Their origin in molecular scattering is conceptually similar to the classic Young's slit experiment where the two atomic centers of a diatomic molecule act as a `molecular' double slit. In practice, the interference pattern depends on the relative magnitudes of the even and odd anisotropies of the PES. An illustrative example of the quantum nature of these propensities can be found in the work of Faure et al. \cite{faure2016} on the rotational excitation of HC$_3$N by H$_2$. These authors showed that the strong even $\Delta j$ propensity observed in quantum close-coupling calculations was entirely absent in QCT results (see their Fig. 2). Recently, the mixed quantum/classical theory (MQCT) has also provided valuable time-dependent insight into this phenomenon \cite{Imanzi2023}. \\
Additionally, scaled AB-He collisional systems have long been a proxy and less expensive alternative for AB-H$_2$ collisional systems. However, it has been shown that the two systems can differ significantly \cite{bouhafs_collisional_2017, bouhafs_rotational_2017, lanza_near-resonant_2014}, and that separate quantum calculations and experimental studies for AB-H$_2$ are needed.\\
Succeeding in accurately measuring relatively high temperature RET rate coefficients and observing the quantum effects in the CO-H$_2$ system would then be a key tool for benchmarking the quantum calculations and would allow more confidence in the use of high temperature RET rate coefficients where CO in collision with H$_2$ is involved; for instance in probing photodissociation regions using high-$j$ CO emission \cite{Joblin2018}, as well as in nonlocal thermodynamic equilibrium models of the heating/cooling balance and physico-chemical processes in proto-planetary disks \cite{woitke_radiation_2009, thi_radiation_2013}.

\section{Theory}

The same theoretical procedure as in \cite{labiad_absolute_2022} was employed. Rigid-rotor scattering calculations were performed with the \textsc{MOLSCAT} code \cite{molscat_v14} combined with the dedicated CO$-$H$_2$ four-dimensional (4D) $\braket{V_{15}}_{20}$ PES. This latter was obtained by averaging the high accuracy full six-dimensional $V_{15}$ PES of Faure et al. \cite{faure_importance_2016} over the $(\upsilon=2, j=0)$ vibrational wavefunction of CO and the $(\upsilon_2=0, j_2=0)$ vibrational wavefunction of H$_2$. The resulting PES was then fitted with a spherical harmonic expansion, as described in \cite{faure_importance_2016} where full details can be found. In particular, a set of 60 basis functions was employed in the scattering calculations, including anisotropies up to $l_1$=7 and $l_2$=4, where the indices $l_1$ and $l_2$ refer to the angular dependence of the CO and H$_2$ orientation, respectively. We note that an even better 6D PES called $V_{23}$ was published recently \cite{stachowiak2024}. For scattering calculations, however, the differences with $V_{15}$ are negligible \cite{Jankowski_private}. It should also be noted that in our previous work, the rigid-rotor approach for CO in $\upsilon=2$ was shown to be very accurate by comparison to full-dimensional scattering calculations (see Fig.~2 in \cite{labiad_absolute_2022}).

The rotational constants of CO and H$_2$ were fixed at the experimental values $B_2$(CO)=1.8875~cm$^{-1}$ and $B_0$(H$_2$)=59.322~cm$^{-1}$. Convergence of the close-coupling calculations was carefully checked with respect to the basis set sizes and propagation parameters. Thus, the rotational basis set of CO included levels $j=0-20$ and that of H$_2$ $j_2=0, 2$ for para-H$_2$ and $j_2=1, 3$ for ortho-H$_2$. The coupled differential equations were solved using the hybrid modified log-derivative Airy propagator from \texttt{RMIN}=3 to \texttt{RMAX}=200~$a_0$ (at low energy) with a step size always lower than 0.25~$a_0$. Cross sections were computed for collision energies in the range $3.9-2200$~cm$^{-1}$ and state-to-state rate coefficients were deduced at 293~K by integrating the cross sections over a Maxwell-Boltzmann distribution of relative velocities. Since the measurements were performed with {\it normal}-H$_2$ (i.e., a para-H$_2$ to ortho-H$_2$ ratio of 1:3), the nuclear-spin weights of 1/4 and 3/4 were finally applied to the rate coefficients of para- and ortho-H$_2$, respectively.

\section{Experimental Method}
\begin{figure}[htb]
        \includegraphics[width=\columnwidth]{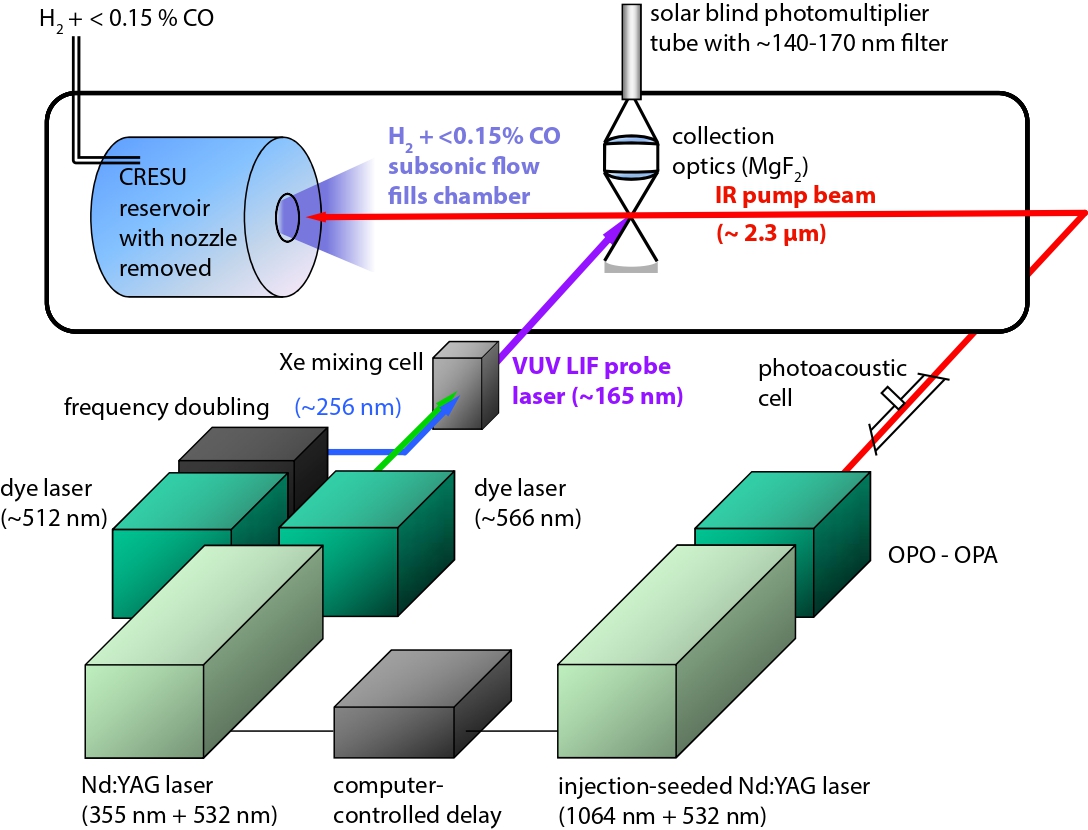}
\caption{Schematic diagram showing an overview of the experimental set up.}
\label{fig:setup}
\end{figure}

The measurements were made in a CRESU (French acronym for Reaction Kinetics in Uniform Supersonic Flow) apparatus combined with the infrared-vacuum ultraviolet, double resonance (IRVUVDR) technique. Although the CRESU apparatus was designed for low temperature studies, the room temperature measurements reported in this paper were not based on supersonic flows, instead, a subsonic laminar flow was generated by removing the nozzle assembly from the CRESU reservoir and introducing the gas in a manner to keep the reservoir and chamber pressures equal, avoiding any gas expansion (see Fig.~\ref{fig:setup}).

CO and H$_2$ gases were introduced and mixed in the reservoir at very low CO concentrations ([CO]/[H$_2$] $<0.15 \%$) in order to minimize CO self-collisions. In the fundamental electronic state, X $^1\Sigma^+$, quantum state-selected rovibrational states ($v = 2$, $j_i$ = 0, 1 or 4) of CO molecules are prepared using tuneable IR laser pumping around 2350~nm. This radiation is generated by a combined optical parametric oscillator – optical parametric amplifier (OPO-OPA, LaserVision) pumped by an injection-seeded Nd:YAG laser (Continuum Surelite EX) operating at a repetition rate of 10~Hz with a pulse-duration of $\sim$5~ns, short enough to probe the fast kinetics of RET. The resulting infrared pulses had an energy of $\sim5-10$~mJ. To ensure high data quality, it is crucial to have sufficiently populated initial rotational states, a condition which is not fulfilled in every IR pulse due to power and frequency fluctuations of the IR laser. A data discrimination process was applied using a photoacoustic reference cell filled with CO, through which the IR laser was passed and fine-tuned to the transition line corresponding to the desired rotational state. Only data points with a photoacoustic signal intensity approximately 1.2 times the average photoacoustic signal were collected.

Under the influence of collisions with H$_2$, initially excited states of CO ($v = 2$, $j_i$) relax to final rovibrational states ($v = 2$, $j_f$). Vibrational relaxation is negligibly slow on the timescale of these experiments. The final states are probed by pulsed laser-induced fluorescence (LIF) via the $\mathrm{A}\,^1\Pi(v=0)\leftarrow\mathrm{X}\,^1\Sigma^+(v=2)$ vacuum ultraviolet (VUV) vibronic band of CO at $\sim$165~nm. This radiation was generated through 4-wave difference frequency mixing in Xe gas as described by Hilbig and Wallenstein \cite{hilbig_tunable_1983}. In this method, two 256.01 nm photons excite the $5p^5 ~6p(5/2)_2 \leftarrow 5p^6~^1S_0$ transition of Xe, which is confirmed by sending the laser to a separate Xe filled reference cell in which a signal of (2+1) Resonance Enhanced Multi-photon Ionisation, REMPI, can be detected. A visible photon was tuned around 566 nm to stimulate the emission of a photon at the same wavelength and generating tuneable VUV at $\sim$165~nm. The 256.01 nm light was produced with a Sirah Lasertechnik Cobra-Stretch dye laser with Coumarin 503 dye (Exciton) pumped by a frequency tripled Continuum Precision II Nd:YAG laser (355 nm) operated at 10 Hz. The output of the dye laser (512.02 nm) was frequency doubled with a BBO crystal to produce $\sim$1~mJ of the desired 256.01 nm light. The frequency doubled output of the Nd:YAG laser (532 nm) also pumped a Laser Analytical Systems LDL 20505 dye laser operated with Rhodamine 6G dye (Exciton) dye to produce $\sim$8~mJ/pulse of $\sim$566~nm light. Both the visible and UV lasers were horizontally polarized, and were focused by an 18 cm (at 256 nm) focal length quartz lens through a fused silica window into a cell filled with 2.7 mbar of Xe. In addition, the visible light was pre-focused through a 150 cm (at $\sim$566~nm) focal length lens. The VUV light produced was re-collimated by a 10.4 cm (at 166 nm) focal length MgF$_2$ lens (Crystran Ltd) into the CRESU chamber perpendicular to the horizontally polarized IR laser beam. In this configuration the double resonance signal is not sensitive to any polarization alignment effects. It is important to perfectly overlap the UV and visible laser pulses inside the Xe mixing cell both in space and time in order to produce the VUV radiation. Following the probe pulse, the resulting LIF was detected by a solar blind photomultiplier tube, (PMT, ET Enterprises 9403B) equipped with a bandpass interference filter (Acton Research) centered at 158.9 nm with a 23.4 nm bandwidth. The signal from the PMT was sent to a Stanford Research SR250 gated integrator with a 16 ns gate and sent to the computer to be recorded. Timing between the IR and the VUV pulses was controlled with two digital delay generators (Stanford Research SR535).
\begin{figure}[htb]
  \includegraphics[width=\linewidth]{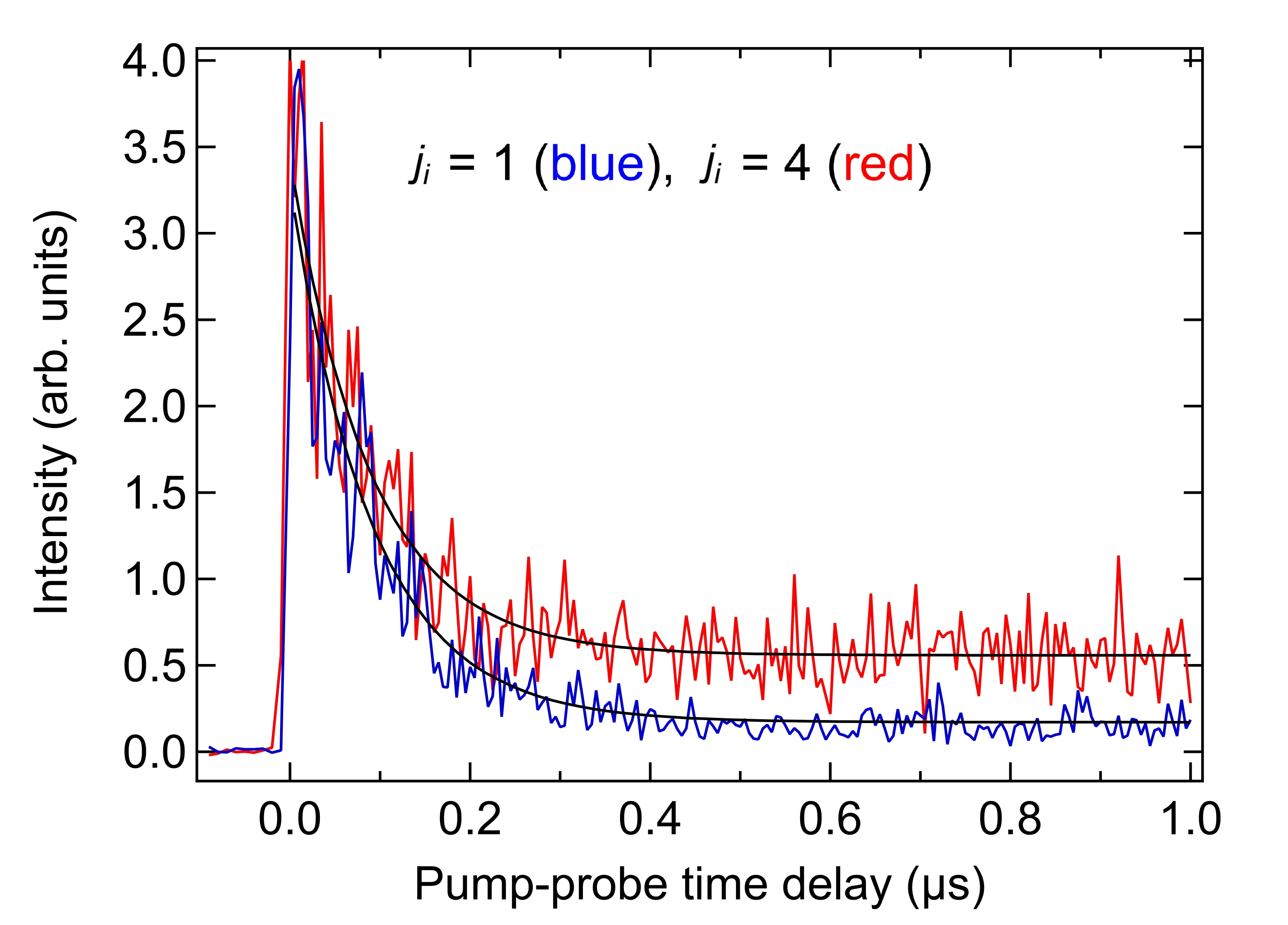}
  \caption {Typical scaled LIF decays (blue and red) for $j_i$ = 1 and $j_i$ = 4, with $[\text{H}_2]$ densities of 1.6 x 10$^{16}$ and 2.4 x 10$^{16}$ molecule cm$^{-3}$, respectively, and their exponential fits (black) used to estimate total removal rate coefficients. The LIF signals are recorded by tuning the pulsed IR laser on the R(0) and R(3) line transitions for state-selective preparation of $j_i$ = 1 and 4 respectively, tuning the pulsed VUV laser on the prepared initial state $j_i$ using the Q(1) and R(4) line transitions for $j_i$ = 1 and 4 respectively and monitoring in time the fluorescence signal using a solar blind PMT.}
  \label{fig:lif}
\end{figure}
\section{Experimental procedure and data analysis} 
\begin{figure*}[htb]
        \includegraphics[width=0.7\linewidth]{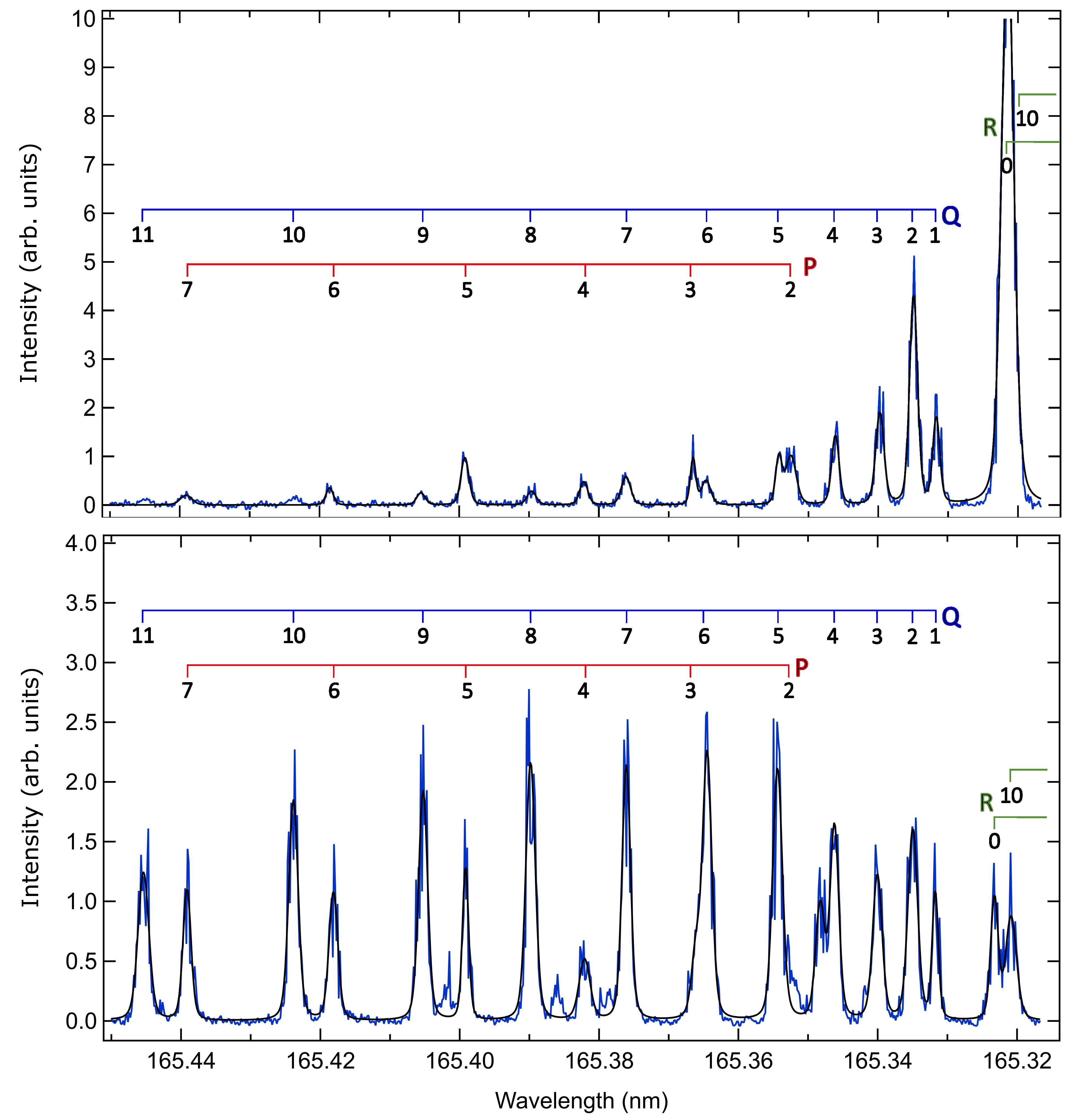}
\caption{Short- and long-time delay spectra for $j_i$ = 0 at 293 K (blue) with multipeak Voigt fitting (black).}
\label{fig:spectra}
\end{figure*}
Similarly to our previous work \cite{labiad_absolute_2022, mertens_rotational_2017}, two types of experiments were performed: a so called \textit{spectroscopic} experiment where state-to-state rate coefficients are measured using spectra recorded at different time delays, and a \textit{kinetic} experiment in which the population decay of the initially pumped rovibrational state is monitored in time. Details of the experimental procedure and data analysis are present in previous papers \cite{carty_rotational_2004,mertens_rotational_2017,labiad_absolute_2022}, and only a short overview is given here.\\

In \textit{kinetic} experiments, total removal rate coefficients are measured by initially preparing a pure single rotational state $j_i$ in $v = 2$ using a pulsed tuneable IR laser, and a second pulsed tuneable VUV laser is used to monitor the LIF decay of the population measured in the line Q($j_i$) as a function of time. Q($j_i$) transition lines of the $\mathrm{A}\,^1\Pi(v=0)\leftarrow\mathrm{X}\,^1\Sigma^+(v=2)$ band of CO are chosen as they are well resolved and more intense compared to other branch lines. Fig.~\ref{fig:lif} illustrates typical time removal signals for $j_i$ = 1 and 4.  First order exponential decay rate coefficients $k_{exp}$ are estimated using a Levenberg-Marquardt fitting algorithm. Reverse collisional processes are considered and assumed to arise from a Boltzmann equilibrium distribution of states in $v = 2$. Total removal coefficients are given by:
 \begin{equation}
   k_{tot} =  \frac{k_{exp}\left(1- f_{j_i}^{eq}\right)}{[\text{H}_2]}
\label{eqn:tot_rates}
\end{equation}
where $f_{j_i}^{eq}$ is the Boltzmann factor of $j_i$ at rotational equilibrium divided by the partition
function, and [H$_2$] is the number density of H$_2$. The LIF signal decays exponentially to a non-zero value where the ratio of this value to the maximum initial intensity represents the Boltzmann factor, which can be used as an additional confirmation of the temperature of the gas, 293 K. 
\begin{figure*}[htb]
        \includegraphics[width=\linewidth]{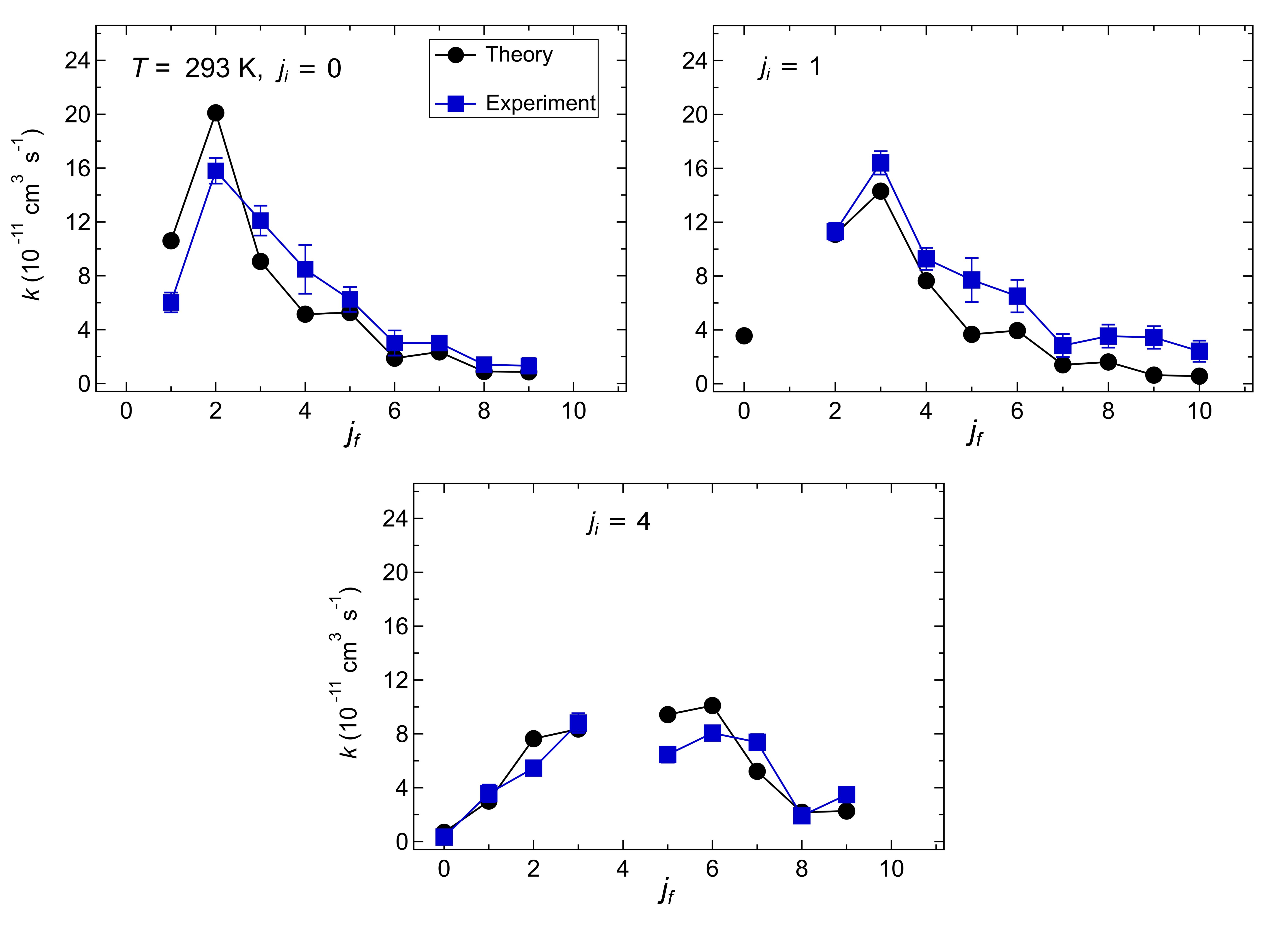}
\caption{Measured state-to-state rate coefficients (squares) and theoretical state-to-state rate coefficients (circles), calculated using the 4D $\langle V_{15} \rangle_{20}$ PES for $j_\mathrm{i}= 0, 1$ and $ 4 \rightarrow j_\mathrm{f}$ at 293\,K. Error bars correspond to 2$\sigma$ statistical errors.}
\label{St-St_RT}
\end{figure*}
In \textit{spectroscopic} experiments, following the initially prepared state $j_i$, a pulsed tunable VUV laser is scanned over all the wavelength range of possibly measured final rotational states $j_f$ ($j_f$ = 0 $-$ 9 or 10). Two spectra are recorded, one at very short time delay when only a small fraction of RET has occurred, typically at 15 to 30 ns after the IR pump, and another one at 2 - 3 $\mu$s when thermal rotational equilibrium has been reached but without significant vibrational relaxation. Voigt lineshapes are fitted to each peak in both spectra (see Fig.~\ref{fig:spectra}) in order to extract the line intensities which are then used to obtain state-to-state rate coefficients according to the formula:\\
\begin{equation}
   k_{j_i \rightarrow j_f} = \left( \frac{I_{j_f}^{\delta t} f_{j_f}^{eq} }{I_{j_f}^{eq}}  \right) \frac{1}{\delta t[\text{H}_2]}
\label{eqn:st-st_rates}
\end{equation}
where $I_{j_f}^{\delta t}$  is the line intensity corresponding to the final rotational state $j_f$ at short time delay $\delta t$, $f_{j_f}^{eq}$ is the Boltzmann factor (fractional population at rotational equilibrium) for $j_f$ and $I_{j_f}^{eq}$ is the line intensity for $j_f$ at equilibrium.
The validity and details of the derivation of equations~(\ref{eqn:tot_rates}) and (\ref{eqn:st-st_rates}) are discussed in previous work \cite{james_combined_1998, carty_rotational_2004}.
\section{Results and discussion}

\begin{table}[htb]
\begin{center}
\begin{tabular}{|c||c|c|c|}
\hline 
 & \multicolumn{3}{c|}{$j_{initial}$} \\ 
\hline 
$j_{final}$ & 0 & 1 & 4  \\ 
\hline 
0 & $j_{initial}$ & - & 0.34 $\pm$ 0.1 (0.7)\\ 

1 & 6.02 $\pm$ 0.7 (10.6)
 & $j_{initial}$ & 3.58 $\pm$ 0.6 (3.0)
 \\ 

2 & 15.8 $\pm$ 0.9 (20.1)
 & 11.3 $\pm$ 0.7 (11.1)
 & 5.46 $\pm$ 0.5 (7.6)
 \\ 

3 & 12.10 $\pm$ 1.1 (9.1)
 & 16.4 $\pm$ 0.9 (14.3)
 & 8.80 $\pm$ 0.7 (8.4)
 \\ 

4 & 8.48 $\pm$ 1.8 (5.2)
 & 9.28 $\pm$ 0.8 (7.7)
 & $J_{initial}$  \\ 

5 & 6.25 $\pm$ 0.9 (5.3)
 & 7.71 $\pm$ 1.6 (3.7)
 & 6.46 $\pm$ 0.6 (9.5)
 \\ 

6 & 3.01 $\pm$ 0.9 (1.9)
 & 6.51 $\pm$ 1.2 (4.0)
 & 8.06 $\pm$ 0.5 (10.1)
 \\ 

7 & 3.01 $\pm$ 0.5 (2.3)
 & 2.84 $\pm$ 0.9 (1.4)
 & 7.38 $\pm$ 0.6 (5.2)
 \\ 

8 & 1.41 $\pm$ 0.5 (0.9)
 & 3.54 $\pm$ 0.9 (1.6)
 & 1.92 $\pm$ 0.5 (2.2)
 \\ 

9 & 1.32 $\pm$ 0.6 (0.9)
 & 3.44 $\pm$ 0.8 (0.7)
 & 3.48 $\pm$ 0.7 (2.3)
 \\ 
 
10 & -
 & 2.42 $\pm$ 0.8 (0.6)
 &  - 

 \\ 
\hline 
$\Sigma k_{st\rightarrow st}$ & 57.4 $\pm$ 2.9 (56.3)
 & 63.5 $\pm$ 3.0 (45.1)
 & 45.5 $\pm$ 1.6 (49.0)
 \\ 

$k_{Total}$ & 32.2 $\pm$ 1.7 (56.3)
 & 42.2 $\pm$ 2.9 (45.1)
 & 33.8 $\pm$ 4.0 (49.0)
 \\ 
\hline 
\end{tabular} 
\end{center}
\caption{Measured state-to-state rate coefficients and theoretical values (in parentheses) for RET in CO -- H$_2$ collisions, calculated using the 4D $\langle V_{15} \rangle_{20}$ PES, for $j_\mathrm{i}=0, 1$, and $4\rightarrow j_\mathrm{f}$ at 293\,K. Units are in 10$^{-11}$\,cm$^3$\,s$^{-1}$ and error limits correspond to 2$\sigma$ statistical errors. The sum of the state-to-state rate coefficients, $\Sigma k,$ and the equivalent rate coefficients, $k_\mathrm{tot}$, from kinetic experiments are also given.}\label{st_293}%

\label{st_table}%
\end{table}%

Table \ref{st_table} and Fig.~\ref{St-St_RT} show very good agreement between the experimental and theoretical state-to-state rate coefficients. The small differences outside of the experimental error bars are likely due to slightly underestimated systematic errors in the measurements. \\
We can make the following observations: 
\begin{enumerate}[(i)]
\item	The rate coefficients decrease with increasing $|$$\Delta j$$|$;
\item	Larger rate coefficients are, in general, observed atroom temperature compared to low temperatures \cite{labiad_absolute_2022}
\item Transitions with $\Delta j$ = +2 are significantly favoured for low $j_i$ (0 and 1);
\item	An even propensity rule is observed for low $\Delta j$, switching to an odd propensity rule for large $\Delta j$;
\item	Different propensity rules and magnitudes are observed compared to the CO-He system.
\end{enumerate}
These observations can be explained as follows. \\
(i) The observed decrease with increasing $|\Delta j| $ is due to the fact that high $\Delta j$ transfers require a greater amount of collision energy according to the general energy gap law \cite{englman_energy_1970}.
We can see though a relatively wide distribution of state-to-state rate coefficients as the CO rotational energy spacing is very small compared to the available collision energy at room temperature, thus favoring large energy transitions.\\
(ii) The generally larger rate coefficients observed at room temperature compared to low temperature values \cite{labiad_absolute_2022} can be explained by two effects causing the increase of RET rate coefficients where each one dominates over a different temperature range: the collisional energy and the long-range 
interaction. The latter dominates at low temperatures as the collisional energy drops below the 88.5 cm$^{-1}$ well depth \cite{faure_importance_2016} i.e. in the resonance regime, and the former dominates at higher temperatures. This is illustrated in Fig.~\ref{fig:Ji_1} where the theoretical and experimental rate coefficients for rotational excitation and de-excitation of CO ($j_i=1$) are plotted as functions of temperature. The resonance regime is observed below $\sim 30$~K for the $j_i=1\to 0$ exothermic transition while the energy criterion applies for all endothermic transitions. We also note that the even $\Delta j$ propensity rule is observed above $\sim 30$~K, due to the larger rotational threshold of the $j_i=1\to 3$ transition (28~K) compared to $j_i=1\to 2$ (11~K). The even $\Delta j$ propensity rule is indeed observed at the cross section level down to the excitation thresholds.
\\
(iii) A clear even $\Delta j$ propensity rule is observed for the CO-H$_2$ system. This is due to quantum interferences which were predicted and explained by McCurdy and Miller \cite{mccurdy_interference_1977} for near homonuclear molecules. Only a few studies have reported experimental observations of even propensity rules \cite{smith_rotational_1982, gijsbertsen_quantum_2006, furio_stateresolved_1986, antonova_state_1997}. To the best of our knowledge, the even propensity rule has never been experimentally observed for CO in collision with any atom or molecular species, except in our previous work at low temperatures where the $\Delta j=+2$ propensity rule was observed for $j_i$ = 0 at 20~K \cite{labiad_absolute_2022}.\\
(iv) In the case of CO in collisions with He, an interference structure was observed at low $\Delta j$ in the experiment of \cite{antonova_state_1999}. However, no clear propensity for even or odd $\Delta j$ was evidenced because their measurements were not sufficiently state-selected. In particular, the initial CO rotational distribution was not state-selective with about 70\% of $j=0$ and 30\% of $j=1$. In this study, we were able to measure pure state-to-state rate coefficients and to observe not only the even propensity for small $\Delta j$, but also the odd propensity for large $\Delta j$, as predicted theoretically by McCurdy and Miller \cite{mccurdy_interference_1977}. As discussed by these authors, resolving this interference structure is a sensitive probe of the anisotropy of the PES because the even and odd propensities directly reflect the competition between terms with even and odd orders in the spherical harmonic expansion of the PES. A change of propensity was thus found for CO + p-H$_2$ by Mengel et al. \citep{Mengel2001} when using the first `high-accuracy' PES for CO-H$_2$ \cite{Jankowski1998}: low temperature rates for $\Delta j=-2$ were found to be smaller than those for $\Delta j=-1$, in contrast to previous studies based on older and less accurate PES. This result, which conflicts with the even propensity rule, was attributed by Mengel et al. to additional resonances in the $|\Delta j=1|$ cross sections calculated with the new PES. An even more significant effect of the PES on the propensity rules was observed in CO + H and was ascribed to the long-range behavior of the PES \cite{Shepler2007, balakrishnan_quantum-mechanical_2002, song_three-dimensional_2013}. It is however important to mention that subsequent refinements of the CO-H$_2$ or CO-H PESs have not changed the propensity rules significantly, while they have noticeably improved the rovibrational spectrum of the complex and the magnitudes of the rate coefficients. We also recall that in this study {\it normal}-H$_2$ was employed so that the magnitude of the rate coefficients is the weighted sum (1/4 and 3/4) of rate coefficients for CO in collision with p-H$_2$ and o-H$_2$. These, however, typically differ by less than 20\%. \\

(v) Compared to the CO-He system \cite{carty_rotational_2004}, CO-H$_2$ exhibits higher rate coefficients and different propensity rule behavior. In particular, rate coefficients for $\Delta j$ = 2 for CO-H$_2$ are four times larger than those of CO-He for $j_i$ = 0 and 1, and double for $j_i$ =  4. This provides clear experimental evidence to underline the now well established observation that AB-He data should not be used as a scaled surrogate for AB-H$_2$ systems. 

\begin{figure}
  \includegraphics[width=\linewidth]{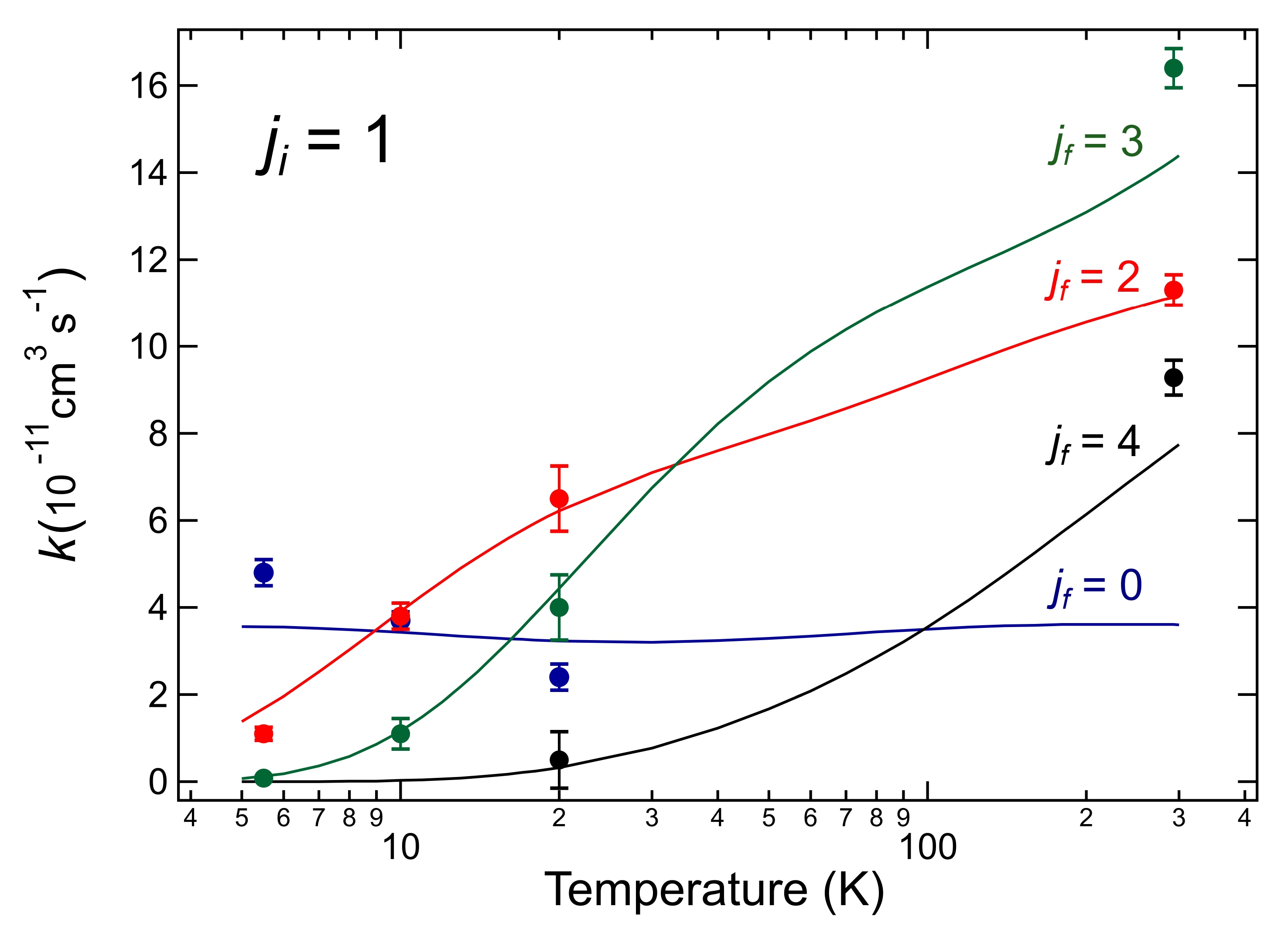}
  \caption{Theoretical (solid lines) and experimental (circles) rate coefficients for the rotational (de)excitation of CO($j_i=4\to j_f=0, 1, 2, 3$) by {\it normal}-H$_2$ as a function of kinetic temperature. Experimental data points between 5.5~K and 20~K are taken from \cite{labiad_absolute_2022}.}
  \label{fig:Ji_1}
\end{figure}
Finally, we note that though the energy gap model does describe the overall dependence of the rate coefficients with $\Delta j$, it is clearly unable to reproduce the detailed behavior observed in this study owing to the predominant even $\Delta j$ quantum propensity rule. 

\section{Summary and conclusion}

Direct absolute state-to-state rate coefficients for RET in the CO-H$_2$ system at room temperature for selected rotational levels $j_i$ in the CO $\mathrm{X}\,^1\Sigma^+$, $v$ = 2 vibrational state were measured using time-resolved IR-VUV double resonance spectroscopy and compared to accurate 4-D quantum scattering calculations. The overall agreement is very good, enabling testing of the anisotropic part of the PES, which is critical for the robust high temperature quantum calculations needed to estimate rate coefficients which can be used to model the CO emission from, for example, photodissociation regions or protoplanetary disks.\\

Experimental evidence is reported for the first time for odd and even propensity rules for the CO-H$_2$ system. This represents a valuable and sensitive tool for benchmarking the anisotropy of the PES. This study also shows that the CO-H$_2$ system behaves in a significantly different way from the CO-He system from the perspective of both the magnitude and general trends of rate coefficients, including the propensity rules. These findings were made possible by the use of the time-resolved IR-VUV double resonance technique which is able to reveal quantum effects even in thermally averaged experiments thanks to high quantum-state selectivity. Finally, in future work, a more detailed theoretical investigation would be valuable (e.g. a partial-wave analysis), for a more comprehensive interpretation of the interference phenomenon in molecular scattering.

\section*{Acknowledgments}
We would like to thank David Carty, Laura Mertens and Martin Fournier for their assistance in the experiments. Thierry Stoecklin, Piotr Jankowski and Krzysztof Szalewicz are acknowledged for long-term collaboration and discussions on the theoretical aspects of this work. All the computations presented in this paper were performed using the GRICAD infrastructure (https://gricad.univ-grenoble-alpes.fr), which is supported by Grenoble research communities.\\
This work was supported by the Agence Nationale de la Recherche (ANR-HYDRIDES), contract ANR-12-BS05-0011-01, by the R\'egion de Bretagne and the French National Programme ``Physique et Chimie du Milieu Interstellaire'' (PCMI) of CNRS/INSU with INC/INP co-funded by CEA and CNES.

\bibliography{CO-H2_RT}
\end{document}